# Highly birefringent polymer terahertz fiber with honeycomb cladding

Yu Hou, Fei Fan, Zi-Wei Jiang, Xiang-Hui Wang, Sheng-jiang Shang[*]

*Institute of Modern Optics, Nankai University, Key Laboratory of Optical Information Science and Technology, Ministry of Education, Tianjin 300071, China*
[*]sjchang@nankai.edu.cn

***Abstract*** **-** Two highly birefringent polymer terahertz (THz) fibers were proposed in this paper, which were formed with honeycomb cladding and some elliptical air holes in the fiber core. The losses and mode birefringence for two different fibers are investigated by finite-different time-domain method. The results show that fiber 2 can achieve both high birefringence (larger than 0.022) and low confinement loss (0.01 dB/m) in a wide THz frequency range. Moreover, compared with a round solid-core fiber, guiding loss of the THz fiber caused by polymer material absorption can be reduced effectively as a part of the mode power is trapped in the air holes.

***Keyword*** **-** Terahertz fiber, birefringence, polymer, honeycomb cladding.

## 1. Introduction

Terahertz (THz) waveguides that can remotely deliver THz radiation have been considered a big step toward compact and robust THz systems and their applications in communication [1], sensing [2], imaging [3], and spectroscopy [4]. At present, most THz systems are bulky and difficult to manipulate because they depend on free space to guide the THz pulse, which requires careful calibration. A flexible waveguide similar to the optical fibers that were developed for shorter wavelength light waves will greatly extend the flexibility and usefulness of THz technology. Within the last few years, various THz waveguides have been proposed and demonstrated [5, 6]. More recently much attention has been paid to birefringent THz fibers, which have potential in applications for polarized THz guiding and filtering. There are three main types of birefringent THz fiber include solid core fibers [7], band gap fibers [8] and porous fibers [9, 10]. These fibers have a common characteristic that can show high birefringence by introducing an asymmetric structure in the fiber core or cladding. According to the report, a polymer terahertz fiber based on a near-tie unit [11] has an extremely large birefringence (5.11 x $10^{-2}$). However, Differences in guidance mechanisms make the fibers have different birefringence and loss. Polarization-maintaining band gap fibers have low confinement loss due to their claddings have air holes arranged periodically while their mode birefringence is usually lower than a level of $10^{-3}$. Compared with the band gap fibers, the porous fibers do not have cladding and the arrangements of air holes in the fiber core are more flexible. So, they have higher birefringence accompanied with poor energy confinement. In this paper, a photonic crystal THz fiber with circular air holes in the honeycomb cladding and elliptical air holes in the fiber core is proposed. According to calculation, both high birefringence (larger than 0.022) and low confinement loss can be achieved in a wide THz frequency range.

First we provide a simple description about our proposed THz fibers. By comparing two different cladding structures of the THz fibers, we show that the changes of the cladding structure have important influence to the THz fiber. Next, the detailed investigation on guided modes and phase-index birefringence of the proposed THz fibers are shown in sectionⅡ. At last, the modal absorption losses and confinement loss are discussed in detail.

## 2. Properties and analysis

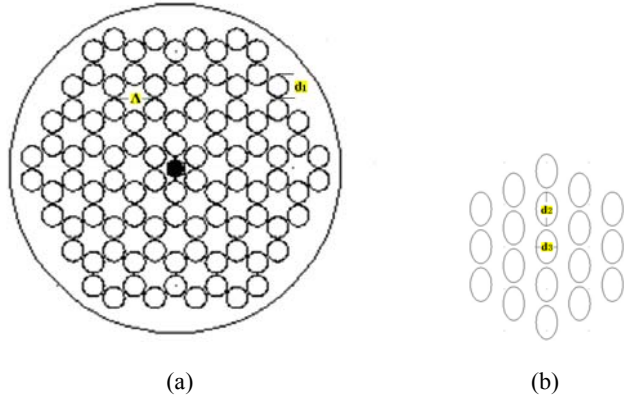

(a)　　　　　　　　　　　　(b)

Fig.1. Cross section of the fiber 1 (a) microstructured core THz fiber with honeycomb cladding (b) the enlarged core area with elliptical air microholes.

Generally THz fibers are typically designed using a tightly packed triangular structure in which the holes are highly inflated to give a high air fill fraction. The high air fill fraction with a different sized air defect in the core is in order to reduce the transmission loss of THz wave. However, this structure is difficult to retain the shape of holes during fabrication due to the holes which have different sizes are arranged too dense, the interaction between the holes is asymmetric, causing the smaller holes to deform. Here we solve the problems by using the honeycomb cladding instead of the triangular structure. The proposed polymer THz fiber is shown in Fig.1. The fiber cladding is consisted of a honeycomb lattice of circular air holes in Topas (a kind of polymer material). The refractive index of Topas is 1.53 and it does not absorb water as opposed to many other polymers. The cladding structure have a hole-to-pitch ratio of $d_1/\Lambda$=0.52 ($\Lambda$=519 μm is a pitch between two holes). In the fiber core some elliptical air micro holes are arranged with a shorter hole pitch，so that the guiding mechanism in such a fiber is based on total internal refection. The length of major axis and minor axis of the elliptical air holes is $d_2$ =24μm and $d_3$ =15μm, respectively.

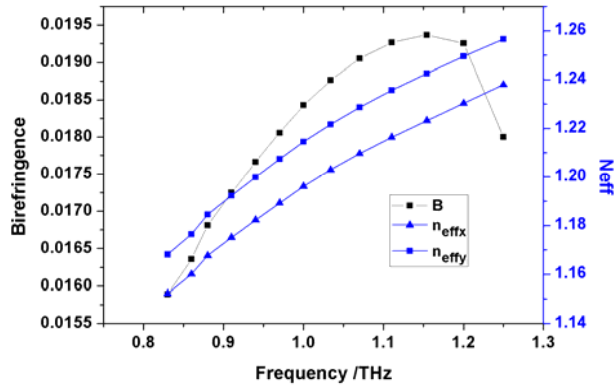

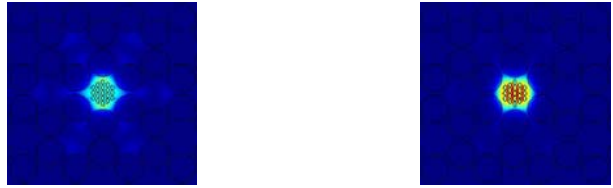

(a)                  (b)

Fig.2. Mode birefringence and effective indices of two polarized modes for the fiber 1. The insets show the time average power flow distribution of *x*-polarized (a) and *y*-polarized (b) modes

The mode birefringence B is defined as $B = n_{effx} - n_{effy}$, where $n_{effx}$ and $n_{effy}$ are the effective refractive indexes of the *x*- and *y*-polarized modes, respectively. As can be seen from Fig.2, we calculate birefringence in a frequency range 0.8 to 1.3 THz for the fiber 1. The mode birefringence increases with frequency initially and reaches its highest value B=0.0192 at f =1.2 THz, Then, the curve decreases quickly because of the x- and y-polarized mode becomes less sensitive to the rise of frequency. Insets a and b in Fig.2 present modal profiles of the fundamental *x*- and *y*-polarized modes at 1 THz, which confined well in the fiber core. But, the details of modal power distributions for the two polarizations are different. We can see that there is more power of the *y*-polarized modes concentrated in the fiber core than that of *x*-polarization modes due to a higher core-cladding index contrast will lead to good confinement. Although the mode birefringence of the fiber reaches a level of $10^{-2}$ over a wide frequency range, we can not reduce large relative absorption loss which is shown in Fig.5. In order to achieve our purpose to obtain a structure with both high birefringence and low loss, we proposed an optimal structure (fiber 2) base on the structure of fiber 1. Some new air holes are added in the cladding and the other structure parameters are not changed.

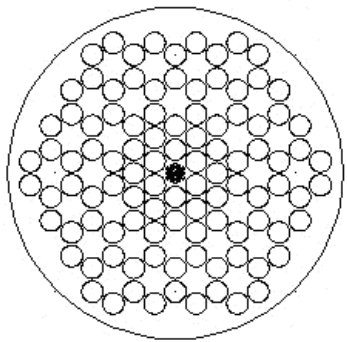

Fig.3. Cross section of the fiber 2 with the same elliptical air microholes in the fiber core

As can be seen from Fig.3, the effective index becomes smaller with the increase of the number of the air hole. The fiber will indicate good confinement of the THz wave due to the core-cladding contrast is improved. We will discuss the loss in the next section. Now, as shown in Fig.4, we calculate the mode birefringence of the fiber 2.

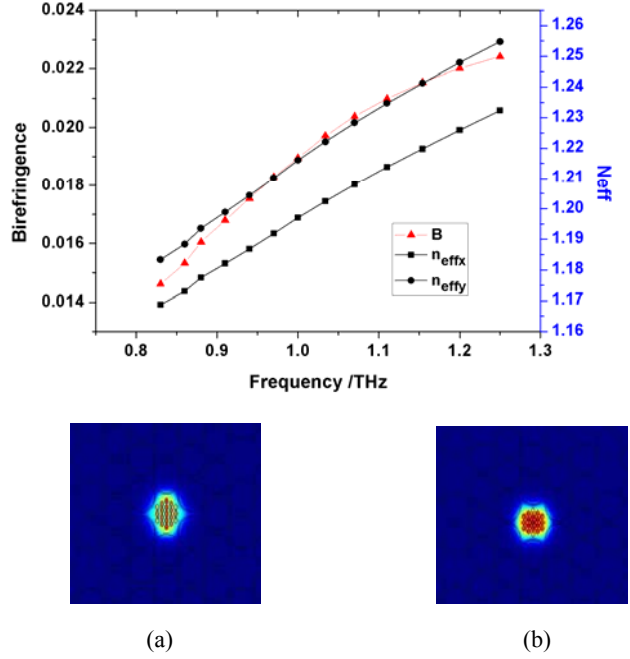

(a)          (b)

Fig.4. Mode birefringence and effective indices of two polarized modes for the fiber 2. The insets show the time average power flow distribution of *x*-polarized (a) and *y*-polarized (b) modes

Compared with Fig.2, we can find that the changes of the fiber cladding have a little effect to the trend of the refractive curve which increases with frequency monotonically. However, it makes the mode birefringence become higher. This is because that the added air holes in the fiber cladding make the effective index of the fiber cladding decrease. The core-cladding contrast causes the fiber core to be more nonsymmetrical. The mode birefringence increases quickly in a frequency range from 0.8 to 1.3 THz which reaches the maximum B = 0.022 at f = 1.25 THz higher than the value of fiber 1. However, if we add more air holes, the restriction on the THz wave and mode birefringence will not change much due to the added air holes are away from the fiber core which suffers from very small impact. Insets of the Fig.4 shows the mode profile of the *x*- and *y*-polarized mode when the incident frequency is 1 THz. Fiber 2 indicates a good confinement of the THz wave. In this case, both high birefringence and low loss will be achieved.

## 3. Loss characteristics of the THz fiber

The power fractions of the mode power in the fiber core which consists of the polymer material and some elliptical holes can be defined as follows:

$$F_x = \frac{\int_x S_z dA}{\int_{all} S_z dA}$$

Where $S_z$ is the time average Poynting vector along the z direction, the subscript x represents the calculated region (polymer material and elliptical air holes), all refers to the region of all the parts. Fig.5 indicates the power fraction of two polarizations in different parts for fiber 1. From the figure, we can find that most of the power is confined in the fiber core which is consistent with our previous analysis. The power fraction of modes in the polymer material increases with frequency monotonically while the power fraction of modes in the elliptical holes deceases slowly. Furthermore, *y*-polarized modes in the polymer materials is higher than that of *x*-polarized modes which shows that the y-polarized modes experience much high loss caused by material absorption.

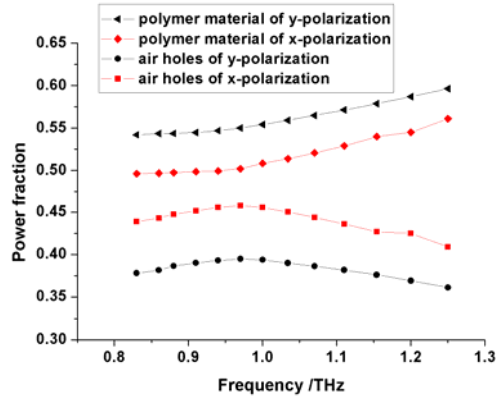

Fig.5. Power fractions of the mode power in polymer material and air core for fiber 1

Fig.6 shows the power fraction of two polarizations in different parts for fiber 2. At low frequencies, most of the modal power distributes in the elliptical holes, indicating that the material absorption loss is low. The power fraction of modes in the elliptical holes deceases as the increase of the frequency while the power fraction of modes in the polymer material rises quickly. They have two intersections at f =1 THz. Compared with fiber 1, we can find that fiber 2 have lower absorption loss and higher mode birefringence.

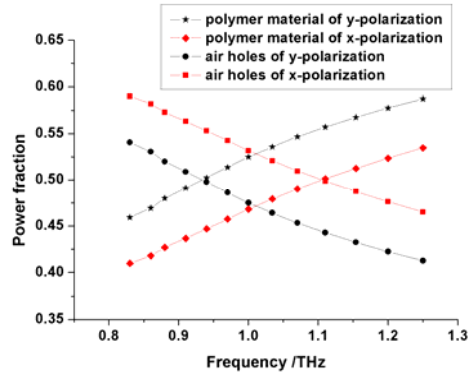

Fig.6. Power fractions of the mode power in polymer material and air core for fiber 2

The different power distributions of the *x*- and *y*-polarized modes result in the different absorption losses of the two fundamental modes. According to a perturbation theory, the absorption loss of our THz fiber can be defined as follows:

$$\frac{\partial_{\mathrm{mod}}}{\partial_{mat}} = \left(\frac{\varepsilon_0}{\mu_0}\right)^{\frac{1}{2}} n \frac{\int_{polymer} |E|^2 \, dA}{\mathrm{Re}\left\{\hat{z} \cdot \int_{all} \vec{E} \times \vec{H}^* \, dA\right\}}$$

Where $n$ is the refractive index of the polymer material. $\partial_{\mathrm{mod}}$ and $\partial_{mat}$ are the absorption coefficients of the fundamental polarization modes and the bulk material. Although $\partial_{mat}$ can be related to the frequency depending on the polymer material, the absorption loss $\partial_{\mathrm{mod}}/\partial_{mat}$ only depends on the parameters of the fiber. Fig.7 indicates $\partial_{\mathrm{mod}}/\partial_{mat}$ for the x- and y-polarized modes of both fibers. We can see that the relative absorption losses increase with the frequency. But, they have different loss values for the same frequency, which shows that the absorption loss of fiber 1 is much higher than that of fiber 2 for the *x*- and *y*-polarized modes. Due to the THz fibers are fabricated with polymer Topas, of which the bulk absorption loss $\partial_{mat}$ is 1 dB/cm at 1 THz [12], the absorption losses for the both polarizations of fiber 1 are 0.55 dB/cm and 0.51 dB/cm, while the losses value of fiber 2 are 0.52 dB/cm and 0.46 dB/cm, respectively. Generally, the $\partial_{\mathrm{mod}}$ of a solid core THz fiber which fabricated by polymer Topas is equal to the bulk absorption loss $\partial_{mat}$ (1 dB/cm at 1 THz). Compare with the THz fiber proposed by us, the material absorption can be reduced effectively as part of the mode power is confined in the air holes.

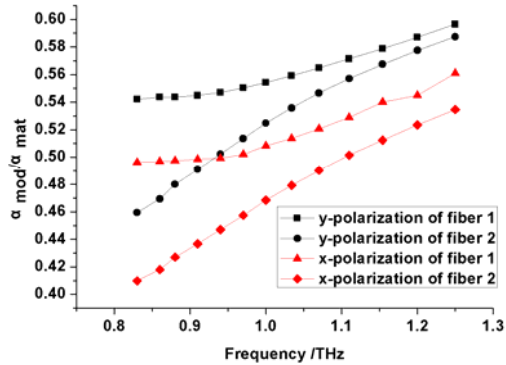

Fig.7. Relative absorption loss of the two fibers for both *x*- and *y*-polarized fundamental modes

The other loss mechanism in fibers is the confinement loss which is closely related to the structure of the cladding. As shown in Fig.8, the confinement losses of both fibers decrease with increased frequency. However, the losses value of fiber 1 are always higher than that of fiber 2 at the same frequency, which indicates that the fiber 2 has a good confinement for THz wave. The confinement loss of the fiber 2 reaches the minimum value (0.01 dB/m) at 1THz, which can achieve both high birefringence (larger than 0.022) and low confinement loss in a wide THz frequency range.

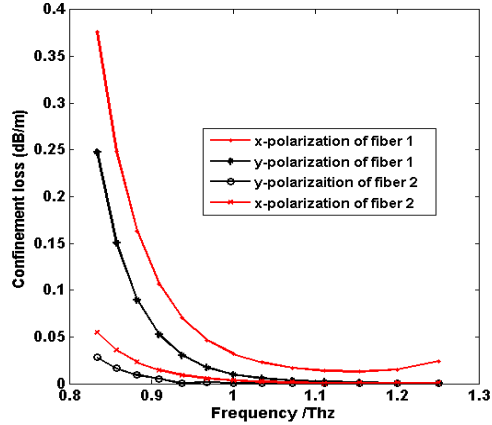

Fig.8. Relative confinement loss of the two fibers for both *x*- and *y*-polarized fundamental modes

## 4. Conclusion

In this paper, we proposed two highly birefringent polymer THz fibers with honeycomb cladding and some elliptical air holes in the fiber core. The losses and mode birefringence for two different structures are investigated and the simulation results show that the fiber 2 can achieve both high birefringence (larger than 0.022 at 1.25 THz) and low absorption loss by confining the dominant fraction of modal power in the air holes in a wide THz frequency range.

Even though the study is purely theoretical, we would like to make a brief comment on the fabrication issues concerning the proposed fibers. PCFs have always attracted a strong interest among the researchers since 1996. In fact, the microstructure presence in the optical fiber cross-section has provided enhanced physical performances which have lead to new developments in different application areas. During the last decade, the fabrication techniques of PCFs were mature. There have been some reports about the design and fabrication of an elliptical-hole PCFs [13, 14]. So, the structures of the two fibers proposed by us were feasible.

## Acknowledgments

This work is supported by the National Basic Research Program of China (973) (Grant No. 2007CB310403), the National Natural Science Foundation of China (Grant No. 61171027)，the Natural Science Foundation of Tianjin of China (Grant No. 10JCZDJC15200) and the Doctoral Fund of Ministry of Education of China (Grant No. 20090031110033).